# Agent Communications toward Agentic AI at Edge – A Case Study of the Agent2Agent Protocol

Qiang Duan* and Zhihui Lu†
*Information Sciences & Technology Department, Pennsylvania State University, Abington, PA, USA
†School of Computer Science and Artificial Intelligence, Fudan University, Shanghai, China

*Abstract*—The current evolution of artificial intelligence introduces a paradigm shift toward agentic AI built upon multi-agent systems (MAS). Agent communications serve as a key to effective agent interactions in MAS and thus have a significant impact on the performance of agentic AI applications. The recent research on agent communications has made exciting rapid progress that leads to a variety of protocol designs, among which the Agent2Agent (A2A) protocol is considered the most representative one. Simultaneously, the rise of edge intelligence is expected to enable agentic AI at the network edge. However, the current agent communication protocols are designed without sufficient consideration of the special challenges of edge computing, and their effectiveness in the edge environment is largely unexamined.

In this paper, we attempt to assess the abilities of agent communication technologies to face the challenges of edge computing using the A2A protocol as a representative case. We first discuss the core functionalities of agent communications, present a landscape of agent communication protocols, and identify the main challenges introduced by edge computing. Then, we conduct a case study on the A2A protocol to examine the key technologies leveraged in the protocol for their effectiveness in meeting the requirements of agent communications in edge computing. Based on the insights obtained from this assessment, we identify open issues in the current agent communication technologies and discuss directions for future research to address these issues.

## I. Introduction

The current evolution in artificial intelligence signals a paradigm shift toward agentic AI, where autonomous agents are capable of perceiving, reasoning, and acting in dynamic environments with minimal human intervention. At the heart of this progression lies a trend moving from single-agent systems to multi-agent systems (MAS), enabling decentralized decision-making, coordination, negotiation, and emergent behaviors that surpass the capabilities of individual agents operating in isolation [1]. Multi-agent systems are thus becoming a technical foundation of the emerging agentic AI.

However, the development of MAS at scale introduces a critical infrastructural challenge: the lack of standardized agent communication protocols. Effective inter-agent communication is the cornerstone of any MAS. Without standard communication protocols, the autonomous agents in a large-scale MAS, especially those developed by different organizations or built upon diverse models, struggle to interoperate. This heterogeneity fragments the agent ecosystem and inhibits the development of an open Internet of Agents demanded by agentic AI.

To address this critical issue, research on agent communications has started attracting broad attention from both industry and academia. Exciting rapid progress has been made in this field recently, leading to the designs of a variety of agent protocols [2]. For example, Google published the Agent2Agent (A2A) protocol in April 2025 [3], soon after Anthropic announced the Model Context Protocol (MCP) [4] in November 2024. Other promising ongoing developments in this area include the Agent Network Protocol (ANP) [5] and Agent Communication Protocol (ACP). Among these efforts of agent protocol designs, the A2A protocol is considered the most representative and broadly embraced, with production-level deployments already underway [6].

Simultaneously, the rise of edge intelligence is expected to reshape the realization of agentic AI. Enabled by advancements in lightweight foundation models and distributed large language model (LLM) deployments, edge computing is now gaining the capability of supporting sophisticated agentic AI applications at the network edge. This realization of agentic AI at the edge is vital for real-time decision-making in latency-sensitive environments such as autonomous vehicles, smart manufacturing, and mission-critical IoT systems. To fulfill the promise of agentic AI at the edge, however, MAS must be adapted to the resource-constrained and heterogeneous nature of large-scale dynamic edge computing systems. Developing robust agent communication frameworks that operate effectively in edge environments remains a key technical hurdle.

Despite the aforementioned rapid progress in research and protocol designs for inter-agent communications, represented by the A2A protocol, the effectiveness of current agent communication protocols in facing the challenges of edge computing is yet largely unexamined. A few comprehensive review papers have been published, either for agent communication protocols in general [2], [7], [8], or about the A2A protocol in particular [6]. However, they all focus on the protocol mechanisms, operations, and applications in typical MAS scenarios without considering the uniqueness of edge computing. Although interesting works have been reported on building agentic AI applications by leveraging various agent communication protocols, mainly leveraging the A2A and MCP protocols; for example, [9], [10], [11], they all assume a deployment environment comprising cloud computing and Internet infrastructure, thus lacking consideration of the special challenges of edge computing.

To our best knowledge, this work is the first attempt to assess agent communication protocols in general and the A2A protocol in particular in terms of their abilities to support inter-agent communications in the edge computing environment toward realizing agentic AI at the edge. In this paper, we first discuss the core functionalities of agent communication for supporting MAS, present a landscape of current agent protocol designs, and identify the main challenges to agent communications in edge computing. Then we conduct an assessment of current agent communications technologies in their effectiveness in facing the challenges of edge computing through a case study on the representative A2A protocol. Based on the insights obtained from this assessment, we identify some open issues in the current design of agent communication protocols and discuss possible directions of future research to address these issues.

## II. Agent Communications in Edge Computing

### A. Agent Communications in Multi-Agent Systems

Agent communications enable interactions among autonomous AI agents to support the intricate dynamics in MAS. A layered architecture is presented in Fig. 1, which illustrates an overall framework for deploying MAS upon the edge computing infrastructure. In this framework, the agent communication layer is positioned between the upper-layer multi-agent system and the underlying edge infrastructure. An agent communication protocol is responsible for utilizing the computation and communication resources provided by the edge infrastructure to build an information exchange platform for supporting the interactions among the AI agents in MAS. Therefore, an agent communication protocol has significant impacts on the MAS performance and thus plays a crucial role in realizing agentic AI.

The core functionalities of an agent communication protocol can be generally categorized into two groups that respectively focus on the following two aspects: i) management of the inter-agent communication system and ii) transportation of information exchange between agents.

The functions for system management govern the description, identification, authentication, publication, and discovery of the AI agents involved in the communication system. The agent description function articulates the capabilities and services offered by an agent in a format that is machine-readable and readily comprehensible by other agents. Agent identification and authentication functions assign a unique identity to each agent and rigorously verify each agent's identity. Agent publication and discovery functions ensure each agent's description is accessible to other agents, thus allowing agents to locate other available agents and select the appropriate one(s) to communicate with for accomplishing specific tasks. The system management aspect of an agent communication protocol is also expected to provide the necessary support demanded by the upper-layer MAS management, typically for agent orchestration functionalities such as agent resource allocation and lifecycle management.

The functions for information exchange pertain to the actual transportation of information between agents. Specifically,

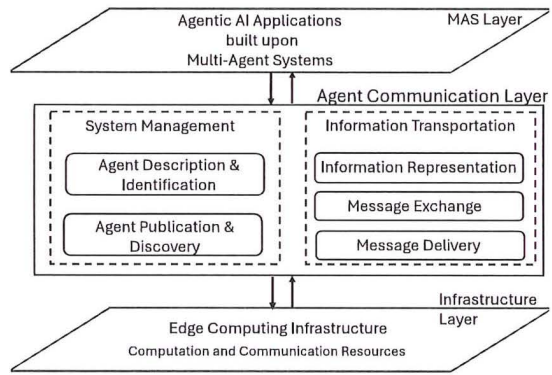

Fig. 1. Agent communication in a multi-agent system

these functions include representation of the exchanged information in specific formats, with options including both structured formats like JSON and unstructured formats such as natural language; patterns of exchanging messages between agents such as synchronous versus asynchronous messaging; and message transportation that ensures the delivery of inter-agent messages across a network, involving sub-functions like sending, forwarding, routing, and receiving messages.

### B. Current Landscape of Agent Communication Protocols

Research on agent communications in MAS has recently gained momentum and led to a wide variety of protocol designs. Although being excited by the rapid progress, researchers who have just entered this emerging field might find the broad spectrum of protocols designed by various organizations focusing on different aspects of the agent communication systems overwhelming or even confusing. In this section, we present a landscape of the current agent protocol developments, with the hope of providing the readers with a big picture that reflects the latest status of this area.

The current agent protocols can be categorized into three groups: i) context-oriented protocols, represented by MCP, are designed for communications between an agent and its context, typically for using tools and/or accessing data sources; ii) user-oriented protocols, for example the Agent-User Interaction protocol (AG-UI) [12], focus on interactions between an agent and front-end applications; and iii) protocols for inter-agent communications in MAS. Protocols in these three categories complement each other and are all needed in a MAS; on the other hand, the inter-agent communication protocols play the key role in realizing agentic AI due to their explicit focus on multi-agent interactions, and thus are the focus of our study in this paper.

There has been a broad spectrum of active research and development on inter-agent communication protocols since early 2025, and exciting progress has been made. The ongoing works in this area can be further divided into two categories: i) general-purpose protocols with a full stack of functions for both system management and information transportation designed for general MAS scenarios, and ii) specific-purpose protocols that either focus on one aspect of inter-agent communications (typically the management aspect) or are tailored to some particular MAS application scenarios.

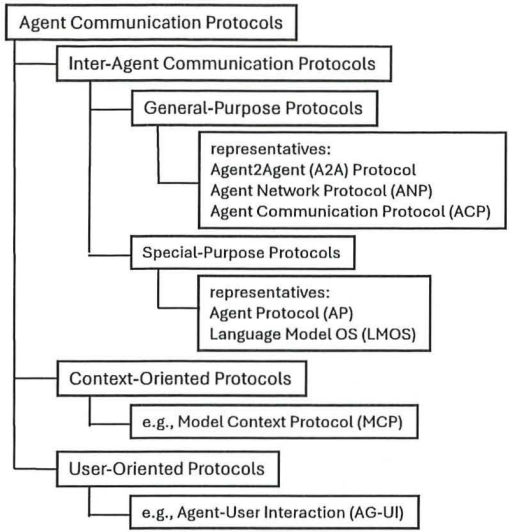

Fig. 2. Categorization of Representative Agent Communication Protocols.

Among general-purpose protocols, the most representative one is the Agent2Agent (A2A) protocol, which was initially published by Google in April 2025 and then transferred to the Linux Foundation as an open standard in June 2025. Other promising developments in this direction include the Agent Network Protocol (ANP), an open-source protocol for cross-domain agent communications, and the Agent Communication Protocol (ACP) developed by IBM using standard BESTful APIs for inter-agent interactions. Representative efforts in the area of special-purpose agent protocols include the Agent Protocol (AP) [13] for agent lifecycle management and the Language Model Operating System (LMOS) [14] focusing on managing and operating agent communications in MAS.

A categorization of agent communication protocols is presented in Fig. 2, which attempts to show a landscape of this rapidly changing field rather than being exhaustive.

Academic research related to agent communication protocols has also been reported recently. For example, Wang *et al.* present a general architecture of the Internet of Agents (IoA) and analyze its key operational enablers in [15]. Chen *et al.* [16] present a framework design with instant-messaging-like information transport and dynamic mechanisms for agent teaming and conversation flow control. The DAWN framework reported in [17] enables distributed agents to be registered, discovered, and organized for building multi-agent communication systems. In [18], the authors design a meta protocol for inter-agent communications to achieve a balance among versatility, efficiency, and portability. Academic research typically focuses on specific mechanisms to address certain aspects of agent communications; thus, complementing the aforementioned protocol designs led by the industry and open-source communities.

### C. Challenges to Agent Communications in Edge Computing

Edge computing involves the deployment of computational resources at the network edge, closer to data sources and end-users, often on devices with limited capabilities. Agentic AI at the edge calls for the deployment of multi-agent systems in the edge computing environment, which introduces challenges to inter-agent communications from the following aspects.

Heterogeneity: An edge-based MAS may comprise agents with highly diverse capabilities and configurations that are deployed on a broad spectrum of edge devices with heterogeneous implementations and various computing/communication capacities. Therefore, edge computing introduces the challenge of heterogeneity both on the agent level and on the host/device level.

Scalability: MAS in edge computing is characterized by the expectation of deploying a vast number of agents across a massive and highly distributed network of edge devices. Managing and coordinating the communications among the numerous, geographically dispersed agents presents significant scalability hurdles.

Dynamicity: There are at least two aspects of dynamicity in the edge computing environment. First, the availability of both computation and communications resources in edge infrastructure keeps varying; for example, edge devices may be switched between on and off to reduce energy consumption, and network bandwidth fluctuates due to changes in wireless channel states. Second, the possible mobility of both mobile devices (as agent users) and edge devices (as agent hosts) introduces extra dynamicity to agent communications in edge computing.

Resource Constraint: The edge devices hosting AI agents often have restricted computing power, storage space, and battery capacities. The network connections among edge devices, which usually rely on wireless mobile communication channels, frequently have constrained bandwidth. On the other hand, the LLMs that form the basis for modern AI agents are particularly resource-greedy and often exceed the capacities of most edge devices.

### III. THE AGENT2AGENT PROTOCOL IN EDGE COMPUTING

Since the A2A protocol is considered representative in protocol design and is the only protocol for inter-agent communications that has been adopted in product-level development of agentic AI applications, we choose A2A as the target of our case study in this section to assess the effectiveness of current agent communication technologies in the edge computing environment toward enabling agentic AI at the edge.

### A. Agent2Agent Protocol Overview

The overall structure of the A2A protocol and its operation process are illustrated in Fig. 3. The A2A protocol involves three actors in agent communications: the *user*, who may be a human or a software application, invokes an agent to fulfill a specific need. A *client agent* translates the user's intent into A2A requests to the remote agent, manages communication sessions with the remote agent, and renders the returned results. A *remote agent* is responsible for processing the task requests received from a client agent and providing responses to the client agent.

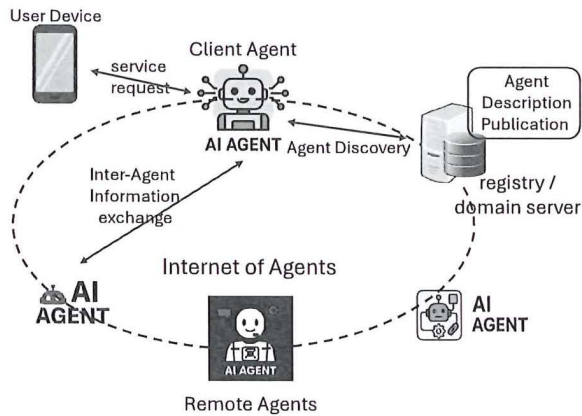

Fig. 3. The overall structure of the Agent2Agent protocol.

The A2A protocol focuses on specifying standard interactions between the client and remote clients, which follow a client-server mode. Upon receiving a user's high-level request, the client agent discovers the available remote agents and selects the appropriate remote agent to which it delegates tasks for accomplishing a goal. Then the client agent initiates a communication session with the remote agent, through which the client and remote agents cooperate to perform the tasks.

The A2A protocol design follows a set of principles. These include embracing agentic capabilities by enabling agents to collaborate in their natural, unstructured modalities; adopting existing standards, including HTTP/HTTPS, JSON-RPC, SSE, etc., to facilitate integration with current IT infrastructure; supporting long-run tasks through mechanisms such as asynchronous messaging, state tracking, real-time feedback, and notification; modality agnostic that supports a wide array of modalities to enable rich multimedia interactions between diverse agents; and opaque execution that allows agent interactions without reveals their internal information, such as reasoning processes, knowledge representations, and tool usages, etc.

#### B. Assessing A2A Protocol for Agent Communications at Edge

In this section, we assess the A2A as a representative protocol for inter-agent communications in MAS for its effectiveness in the edge computing environment. We analyze the protocol's core functionalities, including agent description and identification, agent publication and discovery, and agent information transportation, in terms of their strengths and weaknesses for facing the main challenges of edge computing, including heterogeneity, scalability, dynamicity, and resource constraint.

*1) Agent Description and Identification:* The A2A protocol provides agent description using "Agent Card," which is a structured JSON document that serves as a self-contained description of the agent. An Agent Card typically contains the metadata about an agent, including its name/identifier, a summary of its functions or "skills," the endpoint URL for accessing the agent, and the data formats for required inputs and expected outputs accepted by the agent.

In addition to the standard description of an agent's capabilities and access methods, agent identification and its authentication also play a crucial role in secure agent communications. The A2A protocol assigns unique identities to agents using W3C Decentralized Identifier (DID), which is a globally unique identifier that an agent can create and own without reliance on a central authority. The Agent Card declares the authentication methods supported by the agent, such as OAuth or mTLS.

The standard, structured, and machine-readable agent descriptions provided through Agent Cards in A2A enable the protocol to address the challenges of heterogeneity, scalability, and dynamicity in edge computing. Agent Card allows a large number of heterogeneous agents with diverse abilities and access methods to be described in a standard way, thus they can be found and dynamically added into a multi-agent system as needed. The decentralized identification and authentication mechanisms of the A2A protocol also offer the flexibility that is essential to large-scale communications among different agents in the dynamic edge environment.

However, the current Agent Card specification lacks a standard schema for describing some critical information, particularly host-related details, including OS platforms, capabilities, and resource availabilities (e.g., CPU power, memory space, network bandwidth) of the edge devices that host AI agents. This limitation in agent description constrains the A2A protocol's ability to face the challenge of device-level heterogeneity. It thus lacks support for resource-aware, performance-oriented agent discovery to find the most appropriate agents that are hosted on capable edge devices for meeting the task requirements, which is particularly crucial to agent communications in the resource-constrained edge computing environment.

*2) Agent Publication and Discovery:* The A2A protocol specifies three types of approaches to publishing Agent Cards and discovering available agents. The open discovery approach allows agents to make Agent Cards accessible at a standard "well-known" path *https://agent-domain/.wellknown/agent.json*, then any client capable of resolving the domain via DNS can fetch the card and discover the agent. In the registry-based approach, Agent Cards are published at a registry, and clients can query the registry via an interface (e.g., a RESTful API) exposed by the registry to discover the list of agents published at the registry. The API-based discovery approach limits agents to offering customized API endpoints that deliver Agent Cards only to authenticated clients.

The different agent publication/discovery schemes specified by the A2A protocol cover a spectrum from open web-style DNS-based lookup, to centrally controlled registry query, and to tightly protected private agent catalog. Each approach strikes a different level of balance between accessibility, governance, and security. The flexibility offered by the A2A protocol in agent publication and discovery enables various clients to tailor agent selection and onboarding to meet diverse task objectives; therefore, facilitating MAS to address the heterogeneity challenge to agent communications in edge computing.

However, the current agent publication and discovery mechanisms specified in the A2A protocol are insufficient to fully address the scalability and dynamicity challenges of edge computing. Both the registry-based and API-based approaches are centralized schemes that may form a bottleneck when a large number of available agents need to be published and discovered in a timely manner by a potentially massive number of clients. Although the decentralized open discovery approach may avoid a bottleneck, it assumes each client knows the right domains where its target agents are located, which is unrealistic in a dynamic edge computing environment, where a large number of agents could be hosted in different domains and potentially migrate across domains due to device mobility.

*3) Agent Information Transportation:* We analyze A2A's information transportation functionalities in terms of their effectiveness in edge computing from the following three aspects: information representations, message exchanging patterns, and message delivery mechanisms.

The A2A protocol employs a structured approach for information representation, primarily utilizing JSON-RPC 2.0 as the standardized data exchange format for requests and responses. The A2A design follows the "modality agnostic" principle, supporting various forms of communication beyond text, such as audio, video, and structured data, which enable rich multimedia interactions and accommodate the diverse capabilities of different agents.

The A2A protocol supports a variety of patterns for exchanging messages between agents, designed to handle both instantaneous and long-running tasks efficiently. In the synchronous request/response model, the client agent sends a request and the remote agent returns a direct response. A2A provides two primary mechanisms of asynchronous messaging for long-running processes. Server-Sent Events (SSE) is utilized for real-time streaming of updates from the remote agent back to the client. A client agent also has the option to register webhooks to receive asynchronous push notifications whenever the task's status changes. Furthermore, A2A supports stateful message exchanging by maintaining task states, allowing each stage of an interaction to be correlated to the correct task.

A2A relies on widely adopted web standards for the delivery of inter-agent messages across a network. The primary message transport protocol is HTTP/HTTPS, with production implementations mandating HTTPS with modern TLS ciphers to ensure secure data transmission.

The standardized, structured, and multi-modal approach of information representation in the A2A protocol ensures consistency and flexibility in information transportation across heterogeneous agents. The flexibility in message exchange patterns provided by A2A is vital for adapting to various scenarios and requirements for inter-agent communications, especially in a highly dynamic environment such as mobile edge computing. The A2A protocol leverages "common web standards" for message transportation, thus providing broad compatibility and easing adoption of the protocol across systems with diverse implementations. Therefore, these aspects of the A2A protocol design form a foundation to handle the challenges of heterogeneity and dynamicity brought in by edge computing.

TABLE I
EFFECTIVENESS OF A2A MECHANISMS IN FACING EDGE CHALLENGES
H: HETEROGENEITY, S: SCALABILITY, D: DYNAMICITY, R: RESOURCE CONSTRAINT

| Agent Communication Technologies leveraged in the A2A protocol | Edge Challenges | | | |
|---|---|---|---|---|
| | H | S | D | R |
| standard agent card description | + | + | + | - |
| decentralized agent identification | + | + | + | - |
| open agent publication/discovery | + | + | - | - |
| registry-based agent publication/discovery | + | - | - | - |
| private API for agent publication/discovery | + | - | - | - |
| multi-modal information representations | + | - | + | - |
| flexible message exchange patterns | + | - | + | - |
| web standard-based message delivery | + | - | + | - |
| point-to-point information transportation | + | - | - | - |

On the other hand, the web standards adopted by the A2A protocol for information transportation (e.g., HTTPS, JSON-RPC, SSE, etc.) were not designed or optimized for the unique and often stringent constraints of edge computing, such as limited computational resources on edge devices, intermittent connectivity among hosts, or restricted bandwidth of wireless channels. Therefore, direct applications of such transport schemes in edge computing could cause overwhelming overheads and low efficiency that limit A2A's ability to support large-scale agent communications in resource-constrained edge computing environments. In addition, the current A2A mechanisms for information transportation follow a point-to-point model in which the client agent communicates directly with a remote agent, which does not scale gracefully to large-scale, complex MAS with many interacting agents.

*4) Summary of A2A Protocol Assessment:* In Table I, we summarize our assessment of the key agent communication technologies leveraged in the A2A protocol in terms of their effectiveness in facing the main challenges introduced by edge computing. The symbols "+" and "-" in this table respectively indicate that the mechanism may facilitate or limit the protocol's abilities to face the corresponding challenges.

Overall, the modality-agnostic nature of A2A design, combined with its opaque execution principle, enables the interoperability between diverse autonomous agents that is critical to meet the heterogeneity challenge to agent communications in edge computing. The A2A protocol offers mechanisms that form a promising foundation to achieve large-scale agent communications in dynamic environments; however, further enhancement is needed to fully address the challenges of scalability and dynamicity introduced by edge computing. Agent communications in resource-constrained edge computing appear to be a challenge that the current A2A protocol is least prepared to face, mainly due to the lack of resource-awareness in system management functionalities and the low resource-efficiency of information transportation mechanisms.

More specific discussions about the gap between the current agent communication technologies represented by the A2A protocol design and the expectation for agent communications in edge computing will be presented in the next section, where we will also point out possible research directions to fill the gap.

## IV. OPEN ISSUES AND FUTURE RESEARCH DIRECTIONS

### A. Optimization for Resource Constraints

Interactions among AI agents, which are often LLM-based, could be both computation- and communication-intensive, requiring substantial computing power, memory space, and network bandwidth. However, the current design of the A2A protocol lacks sufficient resource-awareness in core functionalities, including agent description, discovery, and information transportation. Without direct protocol support for fine-grained resource monitoring, allocation, and dynamic adjustment, agent communications in edge computing could easily exhaust edge device capacities as well as network bandwidth, leading to severe performance degradation.

This open issue calls for future research to address the challenge of resource constraints from both the system management and information transportation aspects. Enhancing the management functions to support resource-aware agent discovery and dynamic resource allocation to inter-agent communications is a research topic worth further investigation. A promising direction is to allow cross-layer cooperation between the agent communication protocol and resource management of the underlying infrastructure as well as agent orchestration in the upper-layer MAS.

Resource-efficient information representations and transportation for inter-agent communications also deserve more thorough study. The lightweight transport protocols that have been investigated in the area of edge intelligence (e.g., [19]) may offer some promising approaches. In addition, exploring the leverage of the emerging goal-oriented semantic communications paradigm [20] to facilitate inter-agent interactions would also be an interesting research topic.

### B. Enhancement in Scalability

Although the A2A protocol has employed a variety of mechanisms for supporting agent communications in large-scale multi-agent systems, further enhancement is still expected to fully address the scalability challenge to agent interactions in edge computing.

From the system management aspect, publishing a potentially massive number of agents and discovering the appropriate agents among the large number of available agents for meeting task requirements in a scalable manner, while maintaining other design objectives such as interoperability and authentication among heterogeneous agents, is an essential research problem that deserves further study. Possible solutions to this problem could draw inspiration from service discovery in peer-to-peer networks, like the distributed hash table (DHT) or gossip protocols, that enable efficient information dissemination on a large scale without central coordinators.

In addition, since the less-efficient information representation and transportation, together with the point-to-point message delivery model in current A2A design, also limit the scalability of agent communications in resource-constrained edge computing, the research on optimization of lightweight inter-agent information exchange, as discussed in the previous subsection, is expected to enhance large-scale agent communications in edge computing. Another possible direction worth exploring is to embrace more flexible multi-to-multi message delivery mechanisms, such as a message broker/router or service bus architecture, to allow more scalable multi-agent communications.

### C. Adaptation to Edge Dynamicity

Although a variety of mechanisms have been leveraged by the A2A design to make the protocol resilient to changing agent availability, varying task durations, and fluctuating network conditions, thus forming a foundation for handling the dynamicity challenge of edge computing, comprehensive support for high device mobility and rapid changes in resource status in the edge infrastructure requires more sophisticated mechanisms.

Therefore, effective protocol design for adaptive agent communications to meet the edge dynamicity challenge calls for more research and development. Specific topics that are worth further investigation in this direction include robust session management and seamless handovers/migration for agents hosted on mobile edge devices, integrating location context and probably also resource availability into agent discovery and transport routing decisions, and mechanisms for supporting proactive state synchronization or checkpointing for mobile agents. Also, the service-oriented architecture that allows elastic service provisioning utilizing a converged edge-network-cloud infrastructure [21] offers a promising architectural perspective worth further exploration.

### D. Holistic Design for Multi-tenant Agent Communications

Our analysis has shown that meeting different challenges to agent communications in edge computing may lead to conflicting design objectives, and the technologies leveraged to achieve various objectives are often intertwined and have subtle impacts on each other. For example, interoperability across heterogeneous agents with diverse capabilities demands rich information representations and sophisticated information exchange mechanisms, which, on the other hand, introduce the complexity and overheads that potentially degrade the system efficiency and scalability in a resource-constrained edge computing environment.

Therefore, an architectural design with a holistic view of the overall objectives across the layers of underlying infrastructure, inter-agent communications, and multi-agent inter-actions in MAS is needed to offer comprehensive solutions to this challenging research problem. Agent protocol designs with such a holistic view allow agent communications to fit into the big picture of agent orchestration in MAS. One perspective of such a holistic view is to embrace the notion of *virtualization* in the agent communication architecture, which allows various multi-agent communication systems to coexist upon the shared underlying edge infrastructure. These *virtual* agent communication systems can be configured with protocol settings that prioritize different objectives, e.g., interoperability and adaptability over scalability and efficiency, to meet the diverse requirements of various multi-agent systems.

TABLE II
POSSIBLE RESEARCH DIRECTIONS FOR FACING EDGE CHALLENGES
H: Heterogeneity, S: Scalability, D: Dynamicity, R: Resource Constraint

| Possible Research Directions / Strategies | H | S | D | R |
|---|---|---|---|---|
| cross-layer agent resource management | | ✓ | ✓ | ✓ |
| resource-aware agent description/discovery | | ✓ | ✓ | ✓ |
| lightweight information transportation | | ✓ | | ✓ |
| goal-oriented semantic communication | | ✓ | ✓ | ✓ |
| peer-to-peer agent information dissemination | | ✓ | ✓ | |
| multi-to-multi message transportation | | ✓ | ✓ | |
| service-oriented system design | ✓ | | ✓ | |
| virtualization-based architecture | ✓ | ✓ | ✓ | ✓ |

Table II lists the aforementioned possible research directions and strategies that are worth exploring and the corresponding edge computing challenges that these strategies mainly focus on addressing.

## V. Conclusions

In this paper, we conducted an assessment of the agent communication technologies about their effectiveness in an edge computing environment through a case study of the A2A protocol. We first discussed the core functionalities required for agent communications in MAS, reviewed the current landscape of agent protocol developments, and identified the main challenges to inter-agent communications in edge computing. Then, we particularly examined the A2A protocol as the most representative protocol designed for agent communications, and analyzed how well the key mechanisms leveraged in the A2A protocol may face the challenges introduced by edge computing. Our analysis indicates that although the current agent communication technologies form a solid foundation to meet the heterogeneity challenge and offer promising mechanisms that facilitate agent interactions in a large-scale dynamic edge environment, several critical aspects of agent protocol design require further enhancement to support inter-agent communications in edge computing effectively. Based on the insights obtained from the analysis, we identified open issues in agent communication technologies and discussed possible directions for future research to solve these issues toward fully realizing agentic AI at the network edge.